\documentclass[lettersize,journal]{IEEEtran}
\usepackage{amsmath,amsfonts}
\usepackage{algorithmic}
\usepackage{algorithm}
\usepackage{array}
\usepackage[caption=false,font=normalsize,labelfont=sf,textfont=sf]{subfig}
\usepackage{textcomp}
\usepackage{stfloats}
\usepackage{url}
\usepackage{verbatim}
\usepackage{graphicx}
\usepackage{cite}
\usepackage{xcolor} %
\usepackage{amsmath}
\usepackage{amssymb}
\usepackage[font=small,labelfont=bf]{caption}
\usepackage{wrapfig}
\usepackage{multicol}
\usepackage{paralist}
\usepackage{arydshln}
\usepackage{mathrsfs} % alternat

\usepackage[normalem]{ulem}%For strikethrough

\title{Spectral characterization and
    performance of SPT-SLIM on-chip
    filterbank spectrometers}
\begin{document}

\markboth{IEEE Transactions on Applied Superconductivity. Accepted pre-print version.}%
{This is the author's accepted version which has not been fully edited and content may change prior to final publication.}
\author{

% Addresses
% Authors
\IEEEauthorblockN{
C.~S.~Benson\IEEEauthorrefmark{3},
K.~Fichman\IEEEauthorrefmark{1}\IEEEauthorrefmark{2},
M.~Adamic\IEEEauthorrefmark{4},
A.~J.~Anderson\IEEEauthorrefmark{5}\IEEEauthorrefmark{2}\IEEEauthorrefmark{6},
P.~S.~Barry\IEEEauthorrefmark{3},
B.~A.~Benson\IEEEauthorrefmark{5}\IEEEauthorrefmark{2}\IEEEauthorrefmark{6},
E.~Brooks\IEEEauthorrefmark{6},
J.~E.~Carlstrom\IEEEauthorrefmark{2}\IEEEauthorrefmark{7}\IEEEauthorrefmark{1}\IEEEauthorrefmark{8}\IEEEauthorrefmark{6},
T.~Cecil\IEEEauthorrefmark{8},
C.~L.~Chang\IEEEauthorrefmark{8}\IEEEauthorrefmark{2}\IEEEauthorrefmark{6},
K.~R.~Dibert\IEEEauthorrefmark{6},
M.~Dobbs\IEEEauthorrefmark{4},
K.~S.~Karkare\IEEEauthorrefmark{9},
G.~K.~Keating\IEEEauthorrefmark{10},
A.~M.~Lapuente\IEEEauthorrefmark{9},
M.~Lisovenko\IEEEauthorrefmark{8},
D.~P.~Marrone\IEEEauthorrefmark{11},
J.~Montgomery\IEEEauthorrefmark{4},
T.~Natoli\IEEEauthorrefmark{2},
Z.~Pan\IEEEauthorrefmark{8}\IEEEauthorrefmark{2}\IEEEauthorrefmark{1},
A.~Rahlin\IEEEauthorrefmark{6}\IEEEauthorrefmark{2},
G.~Robson\IEEEauthorrefmark{3},
M.~Rouble\IEEEauthorrefmark{4},
G.~Smecher\IEEEauthorrefmark{12}\IEEEauthorrefmark{4},
V.~Yefremenko\IEEEauthorrefmark{8},
M.~R.~Young\IEEEauthorrefmark{5}\IEEEauthorrefmark{2},
C.~Yu\IEEEauthorrefmark{8}\IEEEauthorrefmark{2}\IEEEauthorrefmark{6},
J.~A.~Zebrowski\IEEEauthorrefmark{2}\IEEEauthorrefmark{6}\IEEEauthorrefmark{5},
C.~Zhang\IEEEauthorrefmark{13},
}
\IEEEauthorblockA{\IEEEauthorrefmark{1}Department of Physics, University of Chicago, 5640 South Ellis Avenue, Chicago, IL, 60637, USA}
\IEEEauthorblockA{\IEEEauthorrefmark{2}Kavli Institute for Cosmological Physics, University of Chicago, 5640 South Ellis Avenue, Chicago, IL, 60637, USA}
\IEEEauthorblockA{\IEEEauthorrefmark{3}School of Physics and Astronomy, Cardiff University, Cardiff CF24 3YB, United Kingdom}
\IEEEauthorblockA{\IEEEauthorrefmark{4}Department of Physics and Trottier Space Institute, McGill University, 3600 Rue University, Montreal, Quebec H3A 2T8, Canada}
\IEEEauthorblockA{\IEEEauthorrefmark{5}Fermi National Accelerator Laboratory, MS209, P.O. Box 500, Batavia, IL, 60510, USA}
\IEEEauthorblockA{\IEEEauthorrefmark{6}Department of Astronomy and Astrophysics, University of Chicago, 5640 South Ellis Avenue, Chicago, IL, 60637, USA}
\IEEEauthorblockA{\IEEEauthorrefmark{7}Enrico Fermi Institute, University of Chicago, 5640 South Ellis Avenue, Chicago, IL, 60637, USA}
\IEEEauthorblockA{\IEEEauthorrefmark{8}High-Energy Physics Division, Argonne National Laboratory, 9700 South Cass Avenue., Lemont, IL, 60439, USA}
\IEEEauthorblockA{\IEEEauthorrefmark{9}Department of Physics, Boston University, 590 Commonwealth Avenue, Boston, MA, 02215, USA}
\IEEEauthorblockA{\IEEEauthorrefmark{10}Harvard-Smithsonian Center for Astrophysics, 60 Garden Street, Cambridge, MA, 02138, USA}
\IEEEauthorblockA{\IEEEauthorrefmark{11}Steward Observatory and Department of Astronomy, University of Arizona, 933 N. Cherry Ave., Tucson, AZ 85721, USA}
\IEEEauthorblockA{\IEEEauthorrefmark{12}Three-Speed Logic, Inc., Victoria, B.C., V8S 3Z5, Canada}
\IEEEauthorblockA{\IEEEauthorrefmark{13}SLAC National Accelerator Laboratory, 2575 Sand Hill Road, Menlo Park, CA, 94025, USA}

}

\maketitle

\begin{abstract} 
The South Pole Telescope Shirokoff Line Intensity Mapper (SPT-SLIM) experiment is a pathfinder for demonstrating the use of on-chip spectrometers for millimeter (mm) Line Intensity Mapping (LIM). We present spectral bandpass measurements of the SPT-SLIM spectrometer channels made on site using a Fourier Transform Spectrometer during the first deployment of SPT-SLIM in the 2024-2025 Austral summer season. We note the effect of FTS systematics on measurements of on-chip filterbank spectrometer resolutions and demonstrate a technique for measuring the narrow band passes of the SPT-SLIM filterbanks that improves beyond the intrinsic resolution of a Fourier Transform Spectrometer.
\end{abstract}
%SPT-SLIM deployed with a focal plane of 5 operational filterbank spectrometers each comprised of 130 spectral filters terminated in a microwave kinetic inductance detector (MKID)

\section{Introduction}
Line Intensity Mapping (LIM) uses imaging spectroscopy to trace emission lines as a function of redshift, producing three-dimensional spatial/spectral maps of the large scale structure (LSS) in the Universe without needing to spatially resolve individual sources. In comparison to optical-wavelength galaxy surveys, LIM can access larger volumes of space and higher redshifts. LIM in the sub-millimeter/millimeter-wavelength band can provide constraints on cosmology, including expansion history~\cite{kakare2018} and the neutrino properties, effective number of neutrino species and the sum of the neutrino masses~\cite{dizgah2022}. While it has significant scientific potential, sub-millimeter/millimeter (sub-mm/mm) LIM is still maturing. Significant advancements in the sensitivity and scalability of large field-of-view mapping spectrometers are needed for sub-mm/mm LIM experiments to be competitive with and complementary to modern galaxy surveys. %However, the field is new, and current instruments lack the sensitivity to compete with galaxy surveys.\\

The South Pole Telescope Shirokoff Line Intensity Mapper (SPT-SLIM) is a pathfinder LIM experiment leveraging integrated on-chip filterbank spectrometers to place constraints on the CO power spectrum in the 120-180 GHz range from the SPT. SPT-SLIM targets CO(2-1) between redshifts $0.3<z<0.9$, CO(3-2) between $0.8 < z < 1.8$, and CO(4-3) between $1.6<z<2.8$~\cite{kakare2022}. SPT-SLIM deployed in the Austral summer of 2024-2025 on the SPT in the camera position previously occupied by the Event Horizon Telescope (EHT) very-long-baseline interferometry (VLBI) receiver~\cite{ehtcamera}. The SPT is an ideal telescope for mm-wave LIM, as the high altitude, low humidity, and stable weather at its location limit atmospheric noise \cite{carlstrom2011}. Due to limited time on-sky, the first deployment of SPT-SLIM has primarily focused on instrumentation and technology demonstration. Bright calibration sources, such as RCW-38 and the Moon, were prioritized to characterize the instrument. Future deployments will have updated detectors and will target a measurement of CO emission lines for LIM science.

To reconstruct the redshift dimension of the LIM signal, constrain atmospheric contamination, and disentangle interloper spectral features, reliable spectral response measurements of SPT-SLIM's spectral channels are crucial. This work gives an overview of the SPT-SLIM detector design and details the spectral characterization of the SPT-SLIM spectrometers during the first SPT-SLIM deployment. We demonstrate measurements of the central frequencies and spectral bandpasses with suitable accuracy beyond the intrinsic resolution of the on-site Fourier Transform Spectrometer (FTS) via spatial-domain fitting of the FTS measurements~\cite{Barry2014PHD,Hailey2014}.

As a companion to this work, details on the SPT-SLIM {Adiabatic Demagnetization Refrigerator (ADR) used in the deployment are given in Young et al. (2025)~\cite{Matt2025} and the atmospheric calibration of the instrument is outlined in Dibert et al (2025)~\cite{Karia2025}. Additionally, details on the RF-ICE detector readout system can be found in Rouble et al. (2022) \cite{rouble2022}.

\section{SPT-SLIM: A Filterbank Spectrometer}
%\begin{figure}
%\centering
%\includegraphics[width=.8\linewidth]{Figures/annotated_slim_filterbank.png}
%\caption{A microscope image of an SPT-SLIM filterbank. The main features of the filterbank are annotated. Image credit: Argonne National Laboratory.}
%\label{fig:filterbank}
%\end{figure}
\begin{figure}
\centering
\includegraphics[width=\linewidth]{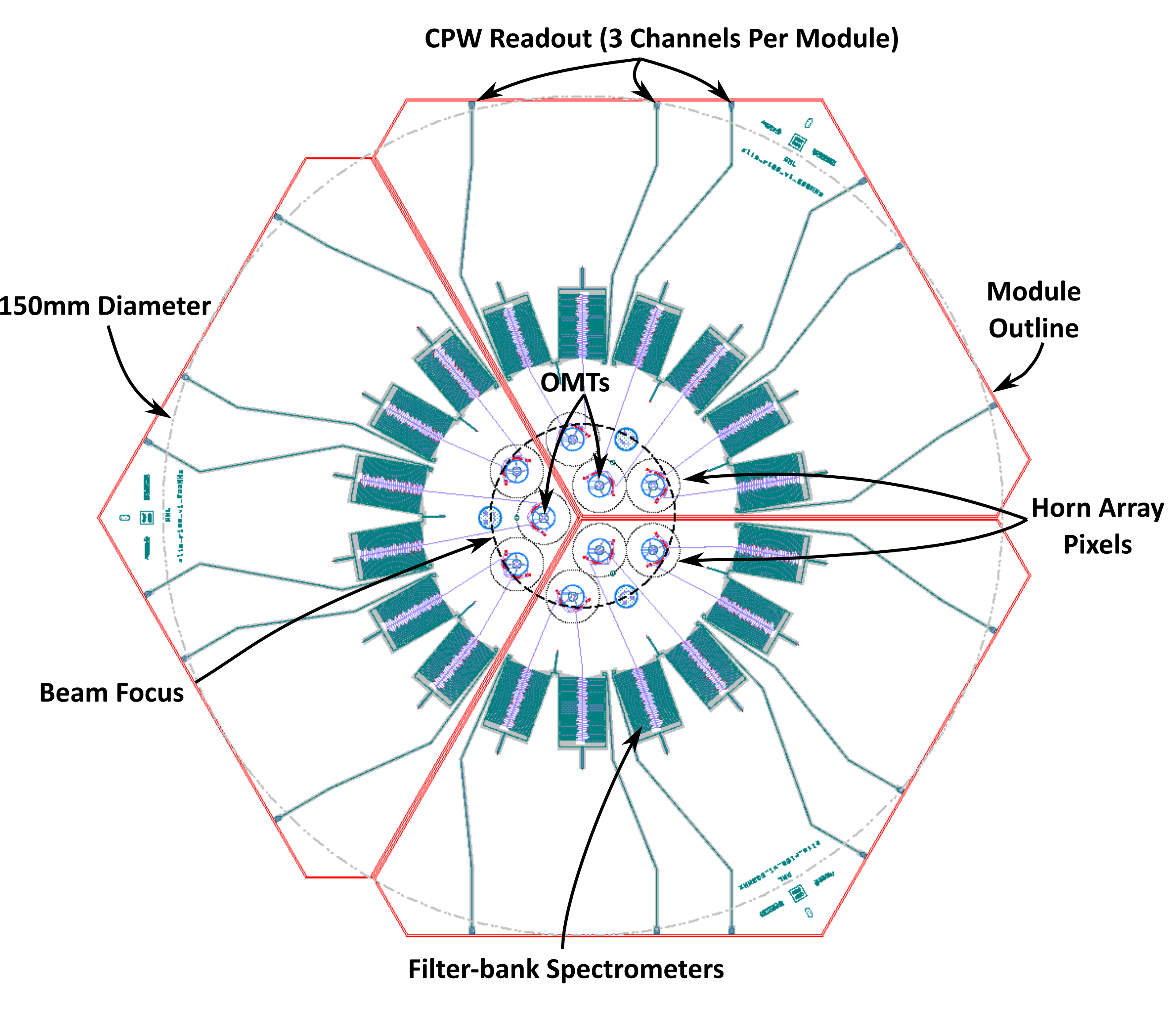}
\caption{A schematic diagram of the full SPT-SLIM focal plane~\cite{robson2024}.}
\label{fig:slimover}
\end{figure}
In the SPT-SLIM design, light is coupled onto polarization-sensitive orthomode transducers (OMTs) via gold-plated conical feedhorns~\cite{Yoon2009,McMahon2009}. The OMTs are patterned on a thin film and mounted on an aluminum base with $\lambda/4$ backshort features. Each polarization mode of the OMT is coupled via a niobium microstrip transmission line to its own filterbank~\cite{barry2022}.

Each of the two filterbanks per antenna consists of 65 half-wave resonator filters capacitively coupled to the transmission line. Each filter provides a narrow spectral bandpass at a set frequency of light, with spectral resolution $R = \nu / \Delta \nu$, that then terminates on a Kinetic Inductance Detector (KID). SPT-SLIM uses a Lumped Element KID (LEKID) design with aluminum inductors and niobium interdigitated capacitors. All LEKIDs within a filterbank are coupled to a single microstrip readout line~\cite{robson2024}. In addition to the narrow-band detectors that build up the spectrometer, each filterbank has three broadband detectors, and two dark detectors to assist in performance characterization. The dark detectors have no optical coupling to the filterbank transmission line.

The SPT-SLIM focal plane consists of three wafers (or ``Submodules''), each with three readout lines, six filterbanks (one per polarization), and three antennae (see Fig.~\ref{fig:slimover}). Due to a combination of damage to the thin film OMTs and non-operational readout lines, five of the nine lines were operational for this first deployment. For additional details on the simulation, design, and fabrication of SPT-SLIM, see Barry et al. (2022)~\cite{barry2022}, Robson (2024)~\cite{robson2024}, Robson et al. (2022)~\cite{Robson2022}, and Cecil et al (2023).~\cite{Cecil2023}.

SPT-SLIM adds to the burgeoning landscape of on-chip filterbank spectrometers providing a multi-pixel on-sky demonstration of the technique. Additionally, SPT-SLIM builds on the design and laboratory testing of SuperSpec~\cite{shirokoff2012} which has been successfully demonstrated through laboratory-based and on-sky measurements in the 255–278 GHz range~\cite{kakare2020}. SPT-SLIM also complements the single-pixel DEep Spectroscopic HIgh-redshift MApper (DESHIMA) experiment, operating on-sky in the 220–440 GHz range~\cite{takekoshi2020}.\\

\section{Measurement}
\begin{figure*}
\centering
  \includegraphics[width=\textwidth]{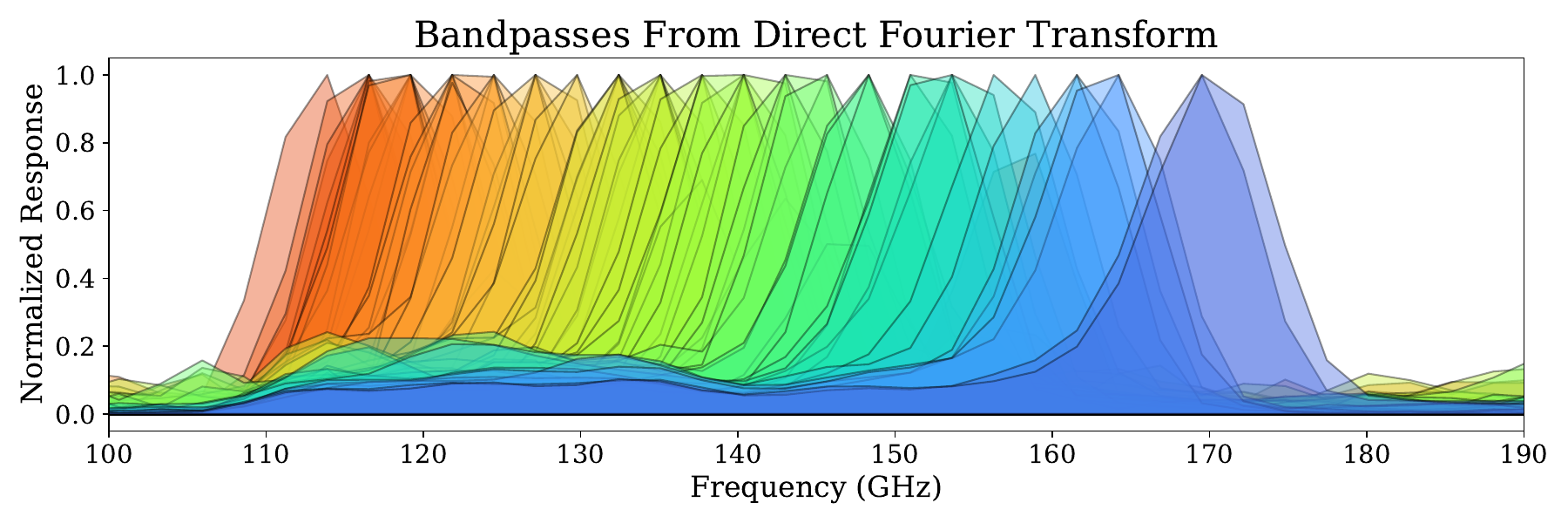}
  \caption{The spectral channels of a polarization-pair of deployment SPT-SLIM filterbanks measured with the FTS. Note that the width of each filter profile has been broadened by the instrumental line shape of the lower-resolution $\Delta \nu\sim 6$\,GHz FTS.}
  \label{fig:slimRainbow}
\end{figure*}
%\begin{minipage}{\textwidth}
    %\includegraphics[width=\linewidth]{Figures/SLIM23_3_spectra_UC.pdf}
   % \includegraphics[width=\linewidth]{Figures/SLIM23_3_spectra_pole.pdf}
    %\vspace{-2mm}
    %\captionof{figure}{Spectra obtained from the Fourier transform of the FTS measurement for the same submodule feedline at University of Chicago (UC) and the South Pole (Pole). The lower resolution of the Pole FTS causes the bandpasses to appear broader at Pole than at UC.}
%\end{minipage}
Characterization of the bandpasses of the SPT-SLIM experiment requires both accurate determination of band centers and bandpass shape. A Fourier Transform Spectrometer (FTS) is an interferometer capable of measuring the spectral response of the filterbank spectrometers with an excellent absolute frequency calibration, with frequency resolution ultimately limited by the size of the FTS}. A measurement is performed by moving a mirror on a motorized platform, varying the differential path length and introducing a phase difference between a stationary arm and the moving arm of the interferometer. The resulting interference signal (i.e., interferogram) measured by the detectors is the Fourier transform of the detector spectra. Here we note that spectroscopic wavenumbers $\sigma=\frac{1}{\lambda}=\frac{\nu}{c}$, where $c$ is the speed of light and $\nu$ is spectral frequency in units of inverse time, provide a natural frequency unit for FTS measurements. The focal plane was measured with the on-site FTS at the South Pole having a minimum resolution element of $d\nu\sim6$\,GHz, corresponding to $R=25$ at 150\,GHz, primarily limited by optical systematics (see~\cite{Pan2019,Pan2020} and the further discussion within this section). Additionally, we note that the minimum resolvable spectral resolution element of an FTS, $\Delta\nu$, is constant across the spectral band. Based on our simulation work, we deemed the on-site FTS sufficient for an initial characterization of SLIM spectrometers. We plan to follow up this work with higher resolution measurements in a laboratory setting.
%reviewer suggestion: Based on our simulation work, we considered the low resolution of the available on-site FTS sufficient for an initial characterization of the SLIM spectrometers. We plan to perform higher-resolution measurements after deployment in a laboratory setting.
%old sentence:Following from the simulation work we outline in this work, we deemed the low resolution of this FTS that was available on site sufficient for an initial characterization of the SLIM spectrometers while following up with a higher-resolution measurement post-deployment in a laboratory setting. 
%SPT-SLIM submodules were given a preliminary partial characterization at the University of Chicago with a higher resolution FTS

The SPT-SLIM spectrometers have a design resolution of $R \sim 100$, significantly greater than that of the FTS at the South Pole. As a result, the spectrometer channel bandpasses are unresolved and broadened by the instrumental line shape of the low-resolution FTS. Bandpasses directly estimated from the as-measured spectrum retrieved by the FTS would provide resolution measurements of the filterbank that is biased to lower resolution (see Fig.~\ref{fig:slimRainbow}). 

We demonstrate that an accurate measurement of an under-resolved filter bandpass can be obtained by relying on the assumed spectral profile of the filters. Using a least-squares fit of the detector signal measured in the spatial displacement domain of the FTS, we extract the filter resolution and central frequency. For this, we have used the functional form of a Fourier-transformed Lorentzian (the corresponding profile of a half-wave filter),
\begin{equation}
    I_0(z) = e^{-\sigma_0 \pi z/R}\cos{(2\pi \sigma_0 z)},
    \label{eq:fitEq}
\end{equation}
where $\sigma_0$ is the central spectral frequency of the bandpass in wavenumbers and $z$ is optical path difference (OPD).

\begin{figure}
\includegraphics[width=\linewidth]{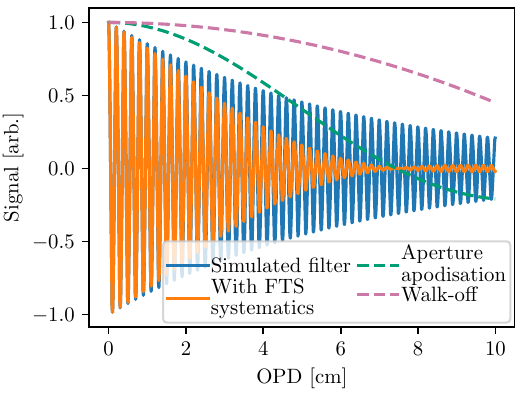}
\caption{The effect of FTS systematics. The dark blue curve shows the simulated interference signal measured by a single SPT-SLIM spectral channel centred at 150\,GHz with an $R=100$ as measured by an ideal FTS that is free of systematics. I.e., the decay of the interference's envelope is exclusively a consequence of the filter's resolution. The dashed lines show the effect of beam walkoff and natural apodization while the orange curve shows the simulated signal of the channel with both of these effects. We note that where the sinc envelope from natural apodiation is negative, a change to the phase of the modulation signal occurrs.}
\label{fig:systematics}
\end{figure}

Coupled with the interference signal from the spectral bandpass of the SPT-SLIM filter, the interferogram of the FTS also experiences natural apodisation and beam walk-off systematics that modify the envelope of the modulation. Walk-off is a result of misalignment between the optical axis of the interferometer and the translational axis of the moving mirror and is well characterized for the FTS at the South Pole in Pan (2020)~\cite{Pan2020}. Natural apodisation of the spectrum is a consequence of the finite entry aperture of the FTS which admits off-axis rays that experience a range of optical path differences as the mirror displaces~\cite{davis2001,locke2009}. As a consequence of these rays, the ideal interferogram measured by the FTS, $I_0(z)$, is multiplied by a sinc envelope given as,
\begin{equation}
    I(z) \approx \Omega\,\mathrm{sinc}\left[\sigma_0 z \left(\frac{\Omega}{2}\right)\right] I_0(z),
    \label{eq:natApod}
\end{equation}
where $\Omega$ is the solid angle of the light entering the FTS (see Fig.~\ref{fig:systematics})\footnote{We note that our initial findings, as presented at this conference and in the initial submission of this manuscript, did not fully account for these FTS systematics, leading to an initial report of systematically low $R$ values.}. Though designed to receive light with a half angle, $\theta$, of 6.3$^\circ$ (see~\cite{Pan2020}) (where $\Omega=\pi\theta^2$), our initial analysis of FTS measurements suggested an incident half-angle of closer to 8$^\circ$, likely due to misalignment and/or some degree of mispositioning between the blackbody source and the FTS. To obtain an estimate of the degree of natural apodisation in our measurements and combat degeneracy with the resolution of the filter, we conducted a simultaneous least-squares fit of selected filter channels to extract a single value for $\Omega$ for each of the spatial pixels of SPT-SLIM (see Equation~\ref{eq:natApod}). These channels were chosen for their high SNR and spectral purity (i.e., quasi-monochromatic). The results of these are shown in Table~\ref{tab:angle}.

\begin{table}
\begin{center}
\caption{The half angle of light entering the FTS, $\theta$, as seen by each OMT. The number of high-quality interferograms fit to determine the angle, $N_{\mathrm{fits}}$, is also given. $^\ast$Though spectrally characterised, pixel 3 of submodule 1 was not used for on-sky measurements.}
\label{tab:angle}
\begin{tabular}{| l | c | c |}
\hline
Channel & $\theta$ [deg] & $N_{\mathrm{fits}}$ \\
\hline
Submodule 1 pixel 1 & 8.3(1)~\, & 16\\
Submodule 1 pixel 2 & 8.1(1)~\, & 17\\
Submodule 1 pixel 3$^\ast$ & 7.7(2)~\, & ~\,5\\
Submodule 2 pixel 1 & 8.0(1)~\, & 12\\
Submodule 2 pixel 2 & 8.13(8) & 37\\
Submodule 2 pixel 3 & 8.1(1)~\, & 21\\
\hline 
\end{tabular}
\end{center}
\end{table}

\begin{figure}
\includegraphics[width=\linewidth]{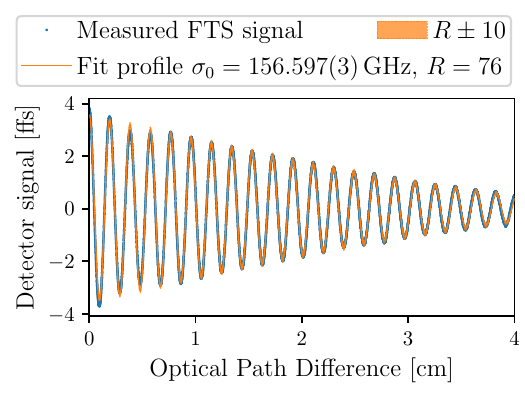}
\includegraphics[width=\linewidth]{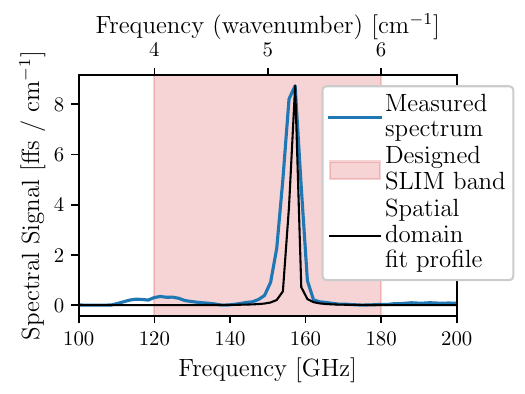}
\caption{A spatial-domain fit of the interference signal from one of the SLIM detectors (top).  The spectrum of the corresponding spectrometer channel measured by the FTS and the spectral profile obtained from the spatial-domain fit are shown in the spectral (Fourier) domain (bottom). Note that the measured profile is smeared to a wider spectral bandpass due to the instrumental line shape of the low-resolution FTS.}
\label{fig:fit_example}
\end{figure}

\begin{figure}
\includegraphics[width=\linewidth]{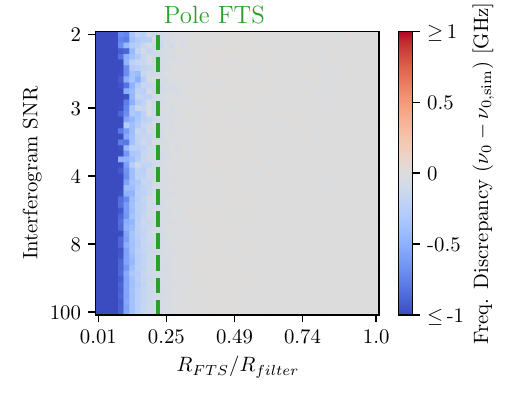}
\includegraphics[width=\linewidth]{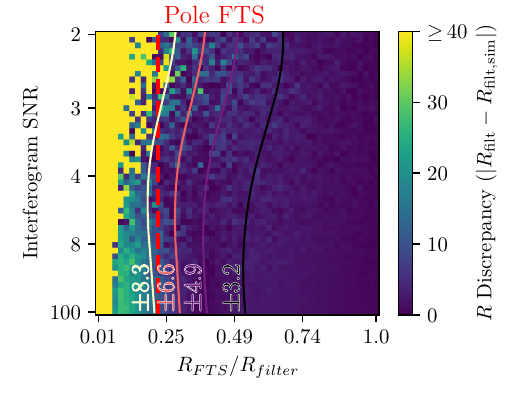}
\caption{The disagreement between simulated values and measurements of the central frequency (top) and spectral resolution/bandpass (bottom) of a filter extracted from spatial-domain fitting using simulated FTS measurements (see Eqs.~\ref{eq:fitEq}--\ref{eq:filterR}). Simulated measurements were generated with a detector noise model at different signal-to-noise ratios (SNRs) and at different FTS resolving powers with each bin/pixel showing the mean of 20 simulated measurements. In each panel, the ratio of the resolving power of the South Pole FTS at 150\,GHz to the $R=100$ simulated filter resolution is marked by a vertical dashed line. The contours in the bottom figure are drawn from the dataset and highlight the increasing uncertainty in filter resolution measurements as the resolution of the measuring FTS decreases and as the SNR of the measurement degrades. For measurements of SNR greater than 3.2 with the South Pole FTS, the uncertainty in values of $R$ extracted from fits of the interferogram is expected to be between 8.3 and 6.6. The uncertainty in measurements of the central frequency of the filter to be less than 0.3\,GHz.}
\label{fig:simulation}
\end{figure}

Fig.~\ref{fig:fit_example} demonstrates fitting the modeled interference pattern corresponding to a single SPT-SLIM filter measured with the low-resolution FTS at the South Pole. The top panel shows that the interference signal is well fit to within the approximate uncertainty of the technique (see Fig.~\ref{fig:simulation}) in the spatial domain. In the Fourier domain, the spectral profile measured directly with the FTS is smeared out by the $\sim$6\, GHz-wide line shape of the FTS while the Fourier transform of the functional form fitted in the spatial domain recovers a narrower, unbiased spectral profile (Fig~\ref{fig:fit_example}, bottom panel).

To determine the accuracy of this technique, the fit was performed on simulated FTS data that incorporates a detector noise model. Model FTS measurements were generated with a variety of noise levels and while truncating the simulated interferogram to different FTS resolutions (note that the resolution of an FTS is determined by the maximum optical path difference measured)\cite{davis2001}. The extracted parameters from the fitting are compared against their simulated values in Fig.~\ref{fig:simulation}. Each bin in the noise and resolution parameter space shown is the the average of 20 simulated spectra. For measurements with any reasonable signal-to-noise ratio (SNR) and relying upon the assumed filter profile, an FTS with an instrumental resolution that is greater than 20\% of the filter resolution is all that is needed to accurately extract the central frequency of the band to within less than 0.2\%. Similarly, for SNRs greater than 8, an FTS with 1/4 the resolution of an $R=100$ filter is sufficient to measure the resolution of the filter to within $\sim$$\pm9$ (as a conservative estimate). Here, we have taken the SNR to be the ratio of the signal amplitude to the root-mean-square of the noise after a high-pass filter has been applied to reduce the modeled 1/$f$ noise of the detectors. We also note that for any reasonable range of spatial sampling rates for the FTS used at the South Pole, providing more than 100$\times$ Nyquist sampling over the range of interference signals in the spectral range of the SLIM band, the accuracy of the fit is unaffected.

Due to the interaction with other filters in the spectrometer, we expect the Lorentzian bandpass of a single filter to experience some asymmetry~\cite{robson2024}. Despite the broadening from the FTS line profile, this can be seen in the measured spectral profile shown in Fig.~\ref{fig:fit_example} (lower panel). To account for this, we introduced a skew to the simulated spectral profiles, $S(\sigma)$, of the form
\begin{equation}
    S_\mathrm{skew}(\sigma) = \left(1+\frac{\alpha}{\tau}(\sigma-\sigma_0)\right)S(\sigma),
    \label{eq:assymProf}
\end{equation}
where $\tau=R/(\pi\sigma_0)$ and $\alpha$ is the introduced skew parameter. For values ranging $|\alpha|=0$--$10\tau$ and measurements with an SNR of 10, we have found that the asymmetry does not significantly affect the values of $\sigma_0$ and $R$ extracted from spatial-domain fits.

%To avoid this bias, the time ordered data were fit with the Fourier transform of a Lorentzian, a decaying sine wave,
%\begin{equation}
 %   I(z) = e^{-sigma_0 \pi z/R}\cos{(2\pi \sigma_0 z)}
%\end{equation}
% to extract values for $R$ and $\sigma_0$, the central frequency of the bandpass. %This method returns an unbiased measurement of $R$.

% \textcolor{red}{discussion of the methodology of the validation simulations}

%\begin{figure}
%    \centering
%    \includegraphics[width=\linewidth]{Figures/Simulation_LTD_Skew.pdf}
%    \caption{The simulated spectral band pass of a filter with and without a modeled asymmetry (see Eq.~\ref{eq:assymProf}).}
%    \label{fig:skew}
%\end{figure}

\section{Frequency Scheduling and Resolution}

\begin{figure*}
\centering
  \includegraphics[width=0.8\textwidth]{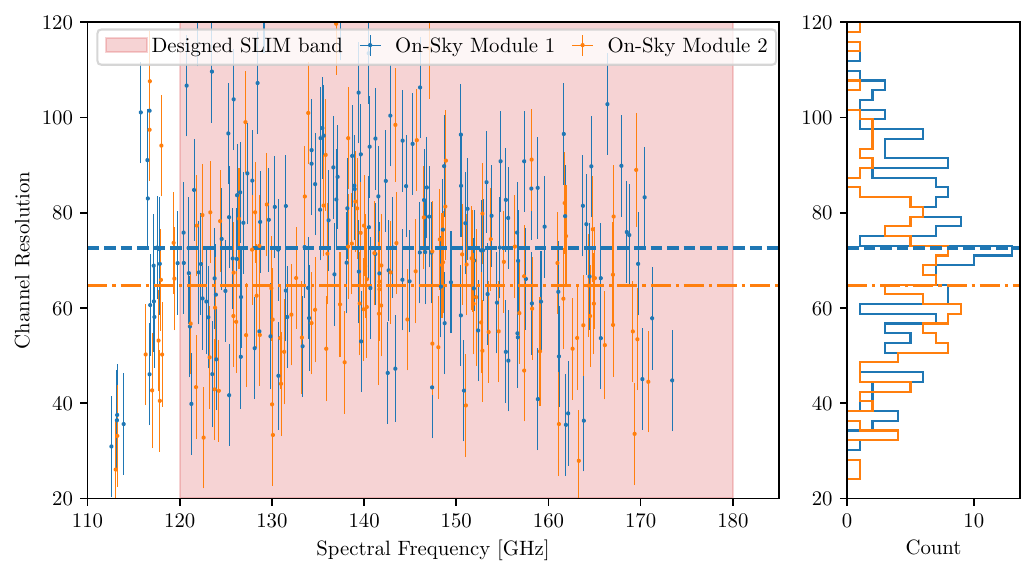}
  \caption{The fitted resolution, $R$, of the on-sky submodules extracted from measurements with the South Pole FTS (see Eq.~\ref{eq:filterR}). Error bars represent the uncertainty determined from the simulated data shown in Fig.~\ref{fig:simulation} at the corresponding SNR of the interferogram measured with the FTS.}
  \label{fig:resolution}
\end{figure*}

Spatial-domain fitting was applied to the FTS measurements of both SPT-SLIM submodules that were operational for its first deployment at the SPT. All three lines on either submodule were measured, although one was not operational for on-sky measurements. These results are shown in Fig.~\ref{fig:resolution}. These show spectral resolution of grouped about median values of 70 and 60 for module 1 and module 2, respectively, each having standard deviations of 30. This is below the design target of $R=100$. Additionally, the centers of the bandpasses are shifted downward from the target SLIM band by $\sim$10 GHz.
\begin{figure}
    \centering
    \includegraphics[width=0.6\linewidth]{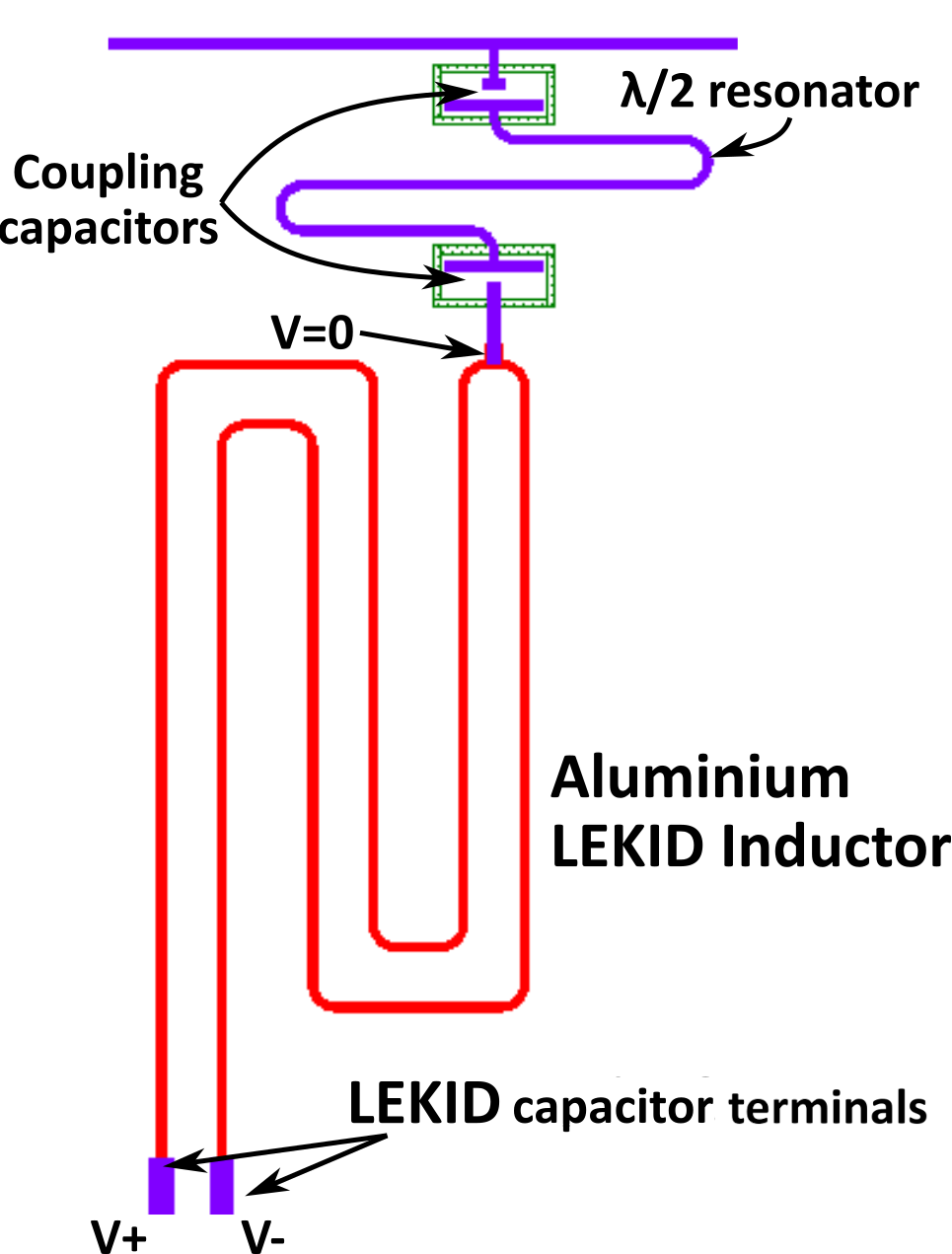}
    \caption{A diagram of a single SPT-SLIM channel. Each filter (i.e., half-wave resonator) is capacitively coupled to the transmission line from the OMT and terminates in a capacitive coupled LEKID. Adapted from Robson (2024)~\cite{robson2024}.}
    \label{fig:filterdiag}
\end{figure}
Each spectral filter consists of a half-wave resonator connected to the transmission line and a detector via coupling capacitors which, together, set the central frequency of the filter’s transmission (see Fig.~\ref{fig:filterdiag}). This was tuned primarily by the physical dimensions of the filter and coupling capacitors and informed by simulations employing projected fabrication errors and an anticipated dielectric loss tangent for the SiN used~\cite{Robson2022}. The shift to lower frequency in the SPT-SLIM bands suggests a systematic deviation from our simulated results that is under study for the next generation of SPT-SLIM spectrometers.

The resolution of a filter is governed by the joined quality factors of the filter's coupling to the feedline, $Q_\mathrm{feed}$, the detector coupling, $Q_t\mathrm{det}$, and losses in the resonator, $Q_\mathrm{loss}$,
\begin{equation}
    \frac{1}{R} = \frac{1}{Q_\mathrm{filt}} = \frac{1}{Q_\mathrm{feed}} + \frac{1}{Q_\mathrm{det}} + \frac{1}{Q_{loss}}.
    \label{eq:filterR}
\end{equation}
In SPT-SLIM, $Q_\mathrm{feed}$ and $Q_\mathrm{det}$ are largely controlled by the dimensions of the coupling capacitors of the half-wave resonator to the feedline and to the detector, respectively. We expect $Q_\mathrm{loss}$ to be dominated by the dielectric loss tangent ($\tan\delta$) of the SiN employed in the microstrip architecture.

The filter resolution measurements shown in Fig.~\ref{fig:resolution} suggest a $\tan\delta$ of $6\,\text{--}\,9\times10^{-3}$, which is within the bounds estimated by experiments characterizing the SiN dielectric at Argonne and optical measurements of test structures on the deployment SLIM submodule wafers~\cite{Pan2023,robson2024}. The shift to lower frequency over the designed spectral band of the SLIM spectral channels may be partly a result of a lower dielectric constant from the value used in the simulations that guided capacitor fabrication (being as much as 15\% lower). Accounting for this possible discrepancy, the $\tan\delta$ may be as low as $\sim5\times10^{-3}$.

%This greatly exceeds what has been measured in characterization experiments of Argonne SiN and optical measurements of test structures on deployment SLIM submodule wafers~\cite{Pan2023,robson2024}. 

The dielectric loss tangent of the SPT-SLIM structures compromises its performance in nearly every aspect, reducing optical efficiency, resolution, and increasing optical loading from the atmospheric emission lines that bookend the SLIM band (particularly when coupled with the overall shift in the frequency bands). %Additionally, shifted central frequencies further exacerbate contamination from the atmospheric lines. 
Lower spectral resolution reduces the resolution in redshift, and therefore reduces potential constraints on cosmological parameters. We expect significant improvements in this respect as we explore alternative low-loss dielectrics while we continue to iterate on our fabrication processes and expand our fabrication potential.

\section{Conclusion}
In the 2024-2025 Austral summer, SPT-SLIM deployed on the SPT in the camera position occupied by the EHT VLBI receiver. This deployment included 5 operational  dual-polarization filterbank spectrometers. These spectrometers were characterized on site with a low-resolution FTS. We demonstrate a technique by which the spectral bandpass/resolution of the filters can be measured with an FTS having less than a quarter the resolution of the designed filter bandwidth. With this, the SPT-SLIM filterbanks deployed were measured to have an average spectral resolution $R\sim65$, lower than the target resolution of $R=100$. This is likely caused by differences between the simulated and fabricated performance of the SiN dielectric. These measured bandpasses are applied to observations of the atmosphere and moon in Dibert et al.~\cite{Karia2025}. \\

\section{Acknowledgments}
This work was supported by funding from the National Science Foundation (NSF), Fermilab National Laboratory under the Department of Energy, and the United Kingdom Research and Innovation (UKRI). SPT-SLIM is supported by the National Science Foundation under Award No. AST-2108763. The South Pole Telescope program is supported by the NSF through awards OPP-1852617 and OPP-2332483. Work at Argonne, including use of the Center for Nanoscale Materials, an Office of Science user facility, was supported by the US Department of Energy, Office of Science, Office of Basic Energy Sciences and Office of High Energy Physics, under Contract No. DE-AC02-06CH11357. This document was prepared by the SPT-SLIM collaboration using the resources of the Fermi National Accelerator Laboratory (Fermilab), a U.S. Department of Energy, Office of Science, Office of High Energy Physics HEP User Facility. Fermilab is managed by FermiForward Discovery Group, LLC, acting under Contract No. 89243024CSC000002. The McGill team acknowledges funding from the Natural Sciences and Engineering Research Council of Canada and the Canadian Institute for Advanced Research, and the Canada Research Chairs program. This work is supported by UKRI Future Leaders Fellowship MR/W006499/1. Partial support is also provided by the Kavli Institute of Cosmological Physics at the University of Chicago. Support for this work for JZ was provided by NASA through the NASA Hubble Fellowship grant HF2-51500 awarded by the Space Telescope Science Institute, which is operated by the Association of Universities for Research in Astronomy, Inc., for NASA, under contract NAS5-26555. KSK was partially supported by an NSF Astronomy and Astrophysics Postdoctoral Fellowship under award AST-2001802, and by SLAC under award LDRD-24-008. We thank Heitor Mourato, Glenn Thayer, and Jose Velho in the BU Scientific Instrument Facility for assistance with mirror fabrication, and John Kovac and Miranda Eiben for assistance with anti-reflection coating optics. 
\bibliographystyle{ieeetr}
\bibliography{references}

 
%\bf{If you include a photo:}\vspace{-33pt}
%\begin{IEEEbiography}
%[{\includegraphics[width=1in,height=1.25in,clip,keepaspectratio]{BiographyPhotos/Chris-Benson-9.jpg}}]{Chris S. Benson.}
%is a research associate at Cardiff University. He received his B.S. (hons.), M.Sc., and Ph.D. degrees in physics and experimental astrophysics from the University of Lethbridge, Canada in 2018, 2020, and 2024, respectively. He has worked on Herschel SPIRE calibration, the development of a Double-Fourier interferometry testbed and the characterisation of its associated detector array, and is working on the spectrometer focal plane of the South Pole Telescope Shirokoff Line Intensity Mapper. His current research interests include on-chip spectrometers, Fourier transform spectroscopy, double-Fourier interferometry, and mm-wave and terahertz detectors.

%[{\includegraphics[width=1in,height=1.25in,clip,keepaspectratio]{New_IEEEtran_how-to.pdf}}]{Michael Shell}
%Use $\backslash${\tt{begin\{IEEEbiography\}}} and then for the 1st argument use $\backslash${\tt{includegraphics}} to declare and link the author photo.
%Use the author name as the 3rd argument followed by the biography text.
%\end{IEEEbiography}



%\vfill

\end{document}